%% file: main.tex
\documentclass[conference,compsoc,a4paper]{IEEEtran}
\IEEEoverridecommandlockouts

\usepackage{cite}
\usepackage{amsmath,amssymb,amsfonts}
\usepackage{algorithmic}
\usepackage{textcomp}
\usepackage{graphicx} 
\usepackage{array} 
\usepackage{booktabs} 
\usepackage{xcolor}
\usepackage{hyperref}
\usepackage{tcolorbox} 
\usepackage{fontawesome5} 
\def\BibTeX{{\rm B\kern-.05em{\sc i\kern-.025em b}\kern-.08em
    T\kern-.1667em\lower.7ex\hbox{E}\kern-.125emX}}

\makeatletter
 \let\old@ps@headings\ps@headings
 \let\old@ps@IEEEtitlepagestyle\ps@IEEEtitlepagestyle
 
 \def\confheader#1{%
  \def\ps@IEEEtitlepagestyle{%
    \old@ps@IEEEtitlepagestyle%
    \def\@oddhead{\strut\hfill#1\hfill\strut}%
    \def\@evenhead{}%
  }%
  \def\ps@headings{%
    \old@ps@headings%
    \def\@oddhead{}%
    \def\@evenhead{}%
  }%
 \ps@headings%
 }

 \makeatother

\confheader{Accepted in 2025 IEEE International Conference on Blockchain and Cryptocurrency (ICBC)}

\usepackage[pscoord]{eso-pic}
\newcommand{\placetextbox}[3]{
 \setbox0=\hbox{#3}
 \AddToShipoutPictureFG*{ \put(\LenToUnit{#1\paperwidth},\LenToUnit{#2\paperheight}){\vtop{{\null}\makebox[0pt][c]{#3}}}
 }
 }
 \placetextbox{.21}{0.055}{\small{979-8-3315-0769-5/25/\$31.00~\copyright 2025 IEEE}}

\begin{document}

\title{Unraveling Ethereum’s Mempool: The Impact of Fee Fairness, Transaction Prioritization, and Consensus Efficiency}

\author{
    \IEEEauthorblockN{S M Mostaq Hossain and Amani Altarawneh}
    \IEEEauthorblockA{\textit{Department of Computer Science}, \textit{Tennessee Technological University}\\
    Cookeville, Tennessee, USA \\
    Email: \{shossain42, aaltarawneh\}@tntech.edu}
}

\maketitle

\begin{abstract}
Ethereum’s transaction pool (mempool) dynamics and fee market efficiency critically affect transaction inclusion, validator workload, and overall network performance. This research empirically analyzes gas price variations, mempool clearance rates, and block finalization times in Ethereum’s proof-of-stake ecosystem using real-time data from Geth and Prysm nodes. We observe that high-fee transactions are consistently prioritized, while low-fee transactions face delays or exclusion—despite EIP-1559’s intended improvements. Mempool congestion remains a key factor in validator efficiency and proposal latency. We provide empirical evidence of persistent fee-based disparities and show that extremely high fees do not always guarantee faster confirmation, revealing inefficiencies in the current fee market. To address these issues, we propose congestion-aware fee adjustments, reserved block slots for low-fee transactions, and improved handling of out-of-gas vulnerabilities. By mitigating prioritization bias and execution inefficiencies, our findings support more equitable transaction inclusion, enhance validator performance, and promote scalability. This work contributes to Ethereum’s long-term decentralization by reducing dependence on high transaction fees for network participation.
\end{abstract}

\begin{IEEEkeywords}
Ethereum Mempool, Transaction Inclusion, Fee Market Efficiency, Consensus Latency, Gas Price Dynamics
\end{IEEEkeywords}

\section{Introduction}
\input{sections/01_introduction}

\section{Research Questions} \label{sec:research_questions}
\input{sections/02_research_questions}

\section{Background} \label{sec:background}
\input{sections/03_background}

\section{Related Works} \label{sec:related_works}
\input{sections/04_related_works}

\section{Methodology} \label{sec:methodology}
\input{sections/05_methodology}

\section{Experimental Setup} \label{sec:experimental_setup}
\input{sections/06_experimental_setup}

\section{Result and Analysis} \label{sec:results_analysis}
\input{sections/07_results_analysis}

\section{Discussion and Improvement} \label{sec:discussion}
\input{sections/08_discussion}

\section{Conclusion} \label{sec:conclusion}
\input{sections/09_conclusion}

\section{Acknowledgments} \label{sec:acknowledgment}
\input{sections/10_acknowledgment}
\bibliographystyle{IEEEtran}
\bibliography{bibfile}

\end{document}

%% file: sections/01_introduction.tex
Ethereum~\cite{buterin2014next}, a leading blockchain platform for smart contracts, relies on the interaction between its execution and consensus layers~\cite{feng2024slimarchive}. The execution layer processes transactions, while the consensus layer ensures agreement on the blockchain’s state. A key component in this system is the transaction pool (mempool)~\cite{wang2024understanding}, where pending transactions compete for inclusion based on gas fees~\cite{meister2024gas}. The efficiency and fairness of this process directly impact transaction inclusion rates, block finalization times, and overall network performance~\cite{grandjean2024ethereum}. To improve fee market efficiency, Ethereum introduced EIP-1559~\cite{buterin2019eip1559}, which replaces the first-price auction model with a dynamic base fee that adjusts per block based on congestion and an optional priority fee (tip) for faster inclusion. The base fee is burned, reducing ETH supply, while the tip incentivizes validators. Understanding these dynamics is essential for optimizing Ethereum’s scalability and transaction processing.


Despite advancements such as EIP-1559, challenges remain in ensuring both efficient and fair transaction inclusion. The mempool plays a crucial role in shaping network congestion~\cite{alipanahloo2024maximum}, as high-fee transactions tend to be prioritized, often leaving lower-fee transactions pending or excluded. This dynamic raises concerns about Ethereum’s fee market fairness and its impact on validator efficiency. While EIP-1559 introduced a more predictable base fee model, fee bidding competition continues to favor high-premium transactions, potentially disadvantaging smaller users~\cite{kiayias2023tiered}. A deeper investigation into how mempool congestion and gas bidding strategies influence network efficiency and fairness is necessary to improve Ethereum’s transaction processing.

Existing research has explored Ethereum's fee structures~\cite{roughgarden2021transaction} and mempool behavior, focusing on gas price volatility and confirmation times. However, a comprehensive study linking mempool congestion to consensus-layer performance, particularly block finalization and validator workload, remains limited. Additionally, while EIP-1559 has been studied for reducing fee volatility~\cite{pierro2019influence}, its impact on transaction fairness is still not fully understood. Addressing these gaps is crucial for enhancing Ethereum's scalability and equitable transaction processing. This research makes the following contributions:
\begin{itemize}
    \item \textbf{Analysis of Mempool-Consensus Interaction:} We investigate how mempool congestion and gas price variations impact block finalization times and validator efficiency.
    \item \textbf{Evaluation of Fee Market Fairness Post-EIP-1559:} While EIP-1559 aimed to stabilize gas fees and reduce volatility~\cite{pierro2019influence}, our findings reveal that fee-based inclusion biases persist and validator efficiency is still constrained during peak congestion. We offer novel data-driven insights that extend beyond prior evaluations of the EIP-1559 mechanism. 
    \item \textbf{Empirical Insights for Protocol Optimization:} Our findings provide data-driven recommendations for improving Ethereum’s fee mechanisms and transaction prioritization.
\end{itemize}
By addressing these issues, this study contributes to a more efficient and fair Ethereum transaction processing framework.
The paper is structured as follows: Section \ref{sec:research_questions} defines key research questions, Section \ref{sec:background} covers Ethereum’s transaction processing and fee mechanisms, and Section \ref{sec:related_works} reviews related literature. Section \ref{sec:methodology} details the research framework, Section \ref{sec:experimental_setup} describes the experimental setup, and Section \ref{sec:results_analysis} presents empirical findings. Section \ref{sec:discussion} interprets results and suggests optimizations, while Section \ref{sec:conclusion} summarizes insights and future research directions.

%% file: sections/02_research_questions.tex
\begin{figure*}[!t]
  \centering
  \includegraphics[width=\linewidth, keepaspectratio]{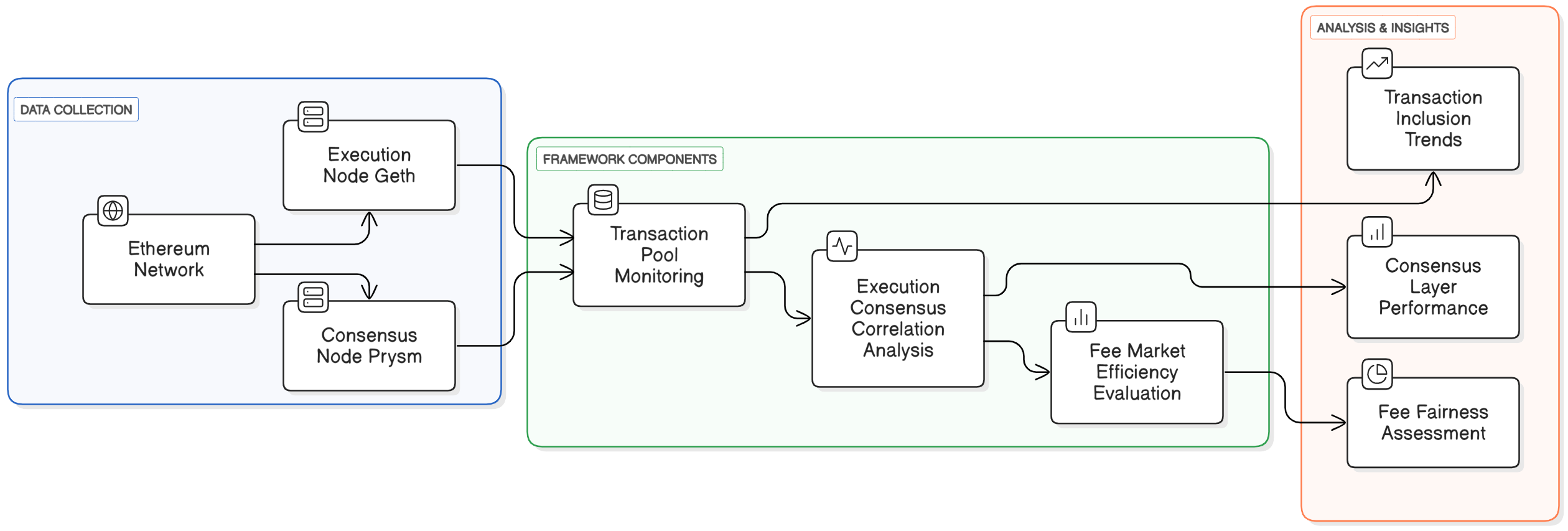}
  \caption{The framework analyzes mempool behavior, execution-consensus correlation, and fee market efficiency using real-time monitoring, statistical modeling, and empirical validation. Key components include transaction pool monitoring, execution-consensus analysis, and fee market evaluation.}
  \label{fig:framework}
\end{figure*}
This work investigates Ethereum’s transaction dynamics, fee fairness, and execution-consensus correlation to identify inefficiencies and biases in transaction processing. By analyzing mempool behavior, fee market fairness post-EIP-1559, and transaction inclusion trends, we aim to provide insights for improving network efficiency and equitable access. The key research questions are:
\begin{itemize}
    \item \textbf{\textit{RQ-i}:} How do Ethereum’s mempool dynamics affect transaction inclusion and network efficiency?
    \item \textbf{\textit{RQ-ii}:} Does Ethereum’s fee market structure post-EIP-1559 ensure fair transaction processing across different fee levels?
    \item \textbf{\textit{RQ-iii}:} How does transaction prioritization impact inclusion fairness, and to what extent do high-fee transactions dominate block space?
\end{itemize}
These questions guide our empirical analysis and contribute to a deeper understanding of Ethereum’s transaction processing mechanisms and potential optimizations.





%% file: sections/03_background.tex
To analyze Ethereum’s transaction dynamics and their impact on execution-consensus efficiency, we first outline key blockchain components: transaction processing, fee mechanisms, mempool operations, and consensus finality.

\subsection{Ethereum Transaction Processing Pipeline}
Ethereum transactions progress~\cite{pacheco2023makes} through multiple stages before finalization. After submission, they enter the mempool, where validators prioritize them based on gas fees and propose blocks for inclusion. The time from submission to finality~\cite{tikhomirov2018ethereum} depends on gas pricing strategies, network congestion, and transaction selection mechanisms. Understanding these processes is crucial for identifying inefficiencies in transaction inclusion and execution delays.

\subsection{Ethereum’s Fee Mechanism and EIP-1559}
Ethereum initially used a first-price auction model for gas fees, resulting in volatile transaction costs~\cite{leonardos2021dynamical}. EIP-1559~\cite{buterin2019eip1559} introduced a dual-fee model with a dynamically adjusted base fee and an optional priority tip. While this change improved fee predictability, concerns persist regarding fee fairness and the prioritization of high-fee transactions, leading to potential exclusion of lower-fee bids. Analyzing Ethereum’s fee structure is essential to understanding its impact on transaction processing.
\subsection{Ethereum Mempool Dynamics}
The mempool acts as Ethereum’s transaction buffer, where pending transactions compete for inclusion based on gas fees~\cite{lyu2022empiricalstudy}. Network congestion and transaction clustering influence clearance rates, while high-fee transactions often bypass lower-fee ones. Investigating how mempool behavior affects validator efficiency and transaction inclusion probability provides insights into network-wide inefficiencies.

\subsection{Consensus Mechanism and Block Finalization}
With Ethereum’s transition to Proof-of-Stake (PoS), validators now propose and attest to blocks~\cite{nguyen2019proof}. However, mempool congestion and transaction prioritization strategies can delay block finalization. Studying the correlation between mempool conditions and consensus efficiency helps optimize validator performance and reduce transaction confirmation delays.

\subsection{Fee Fairness and Transaction Prioritization}
Fairness in Ethereum’s fee market remains a critical issue~\cite{orda2021enforcing}. Our study examines whether the current fee model equitably processes transactions or systematically favors high-fee bidders. By evaluating fee market efficiency and transaction latency, we assess whether Ethereum’s fee structure promotes fair inclusion or exacerbates accessibility disparities.

With these foundational concepts established, the next section reviews related work on Ethereum’s transaction dynamics, fee mechanisms, and consensus behaviors to position our research within the broader blockchain literature.

%% file: sections/04_related_works.tex
Several studies have analyzed the Ethereum mempool, focusing on transaction inclusion policies and gas fee strategies. Research on mempool congestion and transaction prioritization highlights how gas fees influence the speed and likelihood of transaction inclusion~\cite{hausman2024streamlining}. Furthermore, studies on mempool security vulnerabilities emphasize the role of mempool eviction mechanisms, showing that transaction ordering is often manipulated through MEV (Maximal Extractable Value) techniques~\cite{wang2024understanding}. Another study explores alternative lightweight transaction pools for IoT and mobile clients, demonstrating efficiency trade-offs in transaction propagation~\cite{haq2024neonpool}.

Ethereum's EIP-1559 fee model introduced a dynamic base fee~\cite{azouvi2023base} adjustment mechanism to stabilize transaction costs. A research~\cite{leonardos2021dynamical} of the EIP-1559 Ethereum fee market shows that this model lowers fee volatility but doesn't get rid of transaction prioritization biases completely. Another empirical study on Ethereum transaction processing times shows that high gas fees~\cite{meister2024gas} continue to dominate transaction inclusion decisions despite the introduction of base fees~\cite{pacheco2023my}. Furthermore, an investigation~\cite{ocheja2024analytical} into large Ethereum blocks highlights how gas fee auctions and block size constraints affect network congestion and transaction processing speeds.
Correlating Ethereum’s execution and consensus layers is essential to optimizing validator performance. One study shows how validator selection under PoS affects confirmation times~\cite{javed2025empirical}, while another uses machine learning to link gas fee predictions with clearance rates~\cite{oliveira2021analyzing}.
Symbolic fuzzing reveals DoS exploits that manipulate transaction prioritization, raising concerns about mempool resilience~\cite{wang2024understanding}. Fee fairness studies show that low-fee transactions are systematically deprioritized, reducing accessibility for smaller users~\cite{pacheco2023makes}.

Ethereum’s peer-to-peer (P2P) network plays a critical role in transaction propagation and mempool synchronization~\cite{luo2025p2p}. A study on Ethereum’s P2P network structure~\cite{haq2024neonpool} examines client diversity and node discovery inefficiencies, revealing potential bottlenecks that impact transaction broadcasting. Research into Ethereum’s blob fee market~\cite{heimbach2025blob} also sheds light on the complexities of gas price adjustments and their impact on transaction finalization~\cite{pacheco2023makes}. While our study does not directly model MEV extraction techniques, the observed fee-based inclusion biases and congestion patterns offer insights relevant to MEV behavior. As shown in~\cite{alipanahloo2024maximum}, mempool conditions affect transaction ordering and bundling strategies; our findings on prioritization under fee pressure highlight systemic patterns that MEV actors could exploit or be constrained by.

\begin{table}[!t]
\caption{Comparison with Related Work}
\centering
\begin{tabular}{@{}p{1.8cm}p{3cm}p{2.8cm}@{}}
\toprule
\textbf{Study} & \textbf{Focus} & \textbf{Gap Addressed} \\
\midrule
Leonardos et al.~\cite{leonardos2021dynamical} & EIP-1559 modeling & No live-node validation \\
Pacheco et al.~\cite{pacheco2023my} & Transaction timing study & Lacks fairness or validator insights \\
Wang et al.~\cite{wang2024understanding} & Mempool DoS analysis & No latency or fairness view \\
Alipanahloo et al.~\cite{alipanahloo2024maximum} & MEV defense focus & Doesn’t generalize to fee fairness \\
\textbf{This Work} & Fee fairness + validator efficiency & Post-EIP-1559 congestion + inclusion link \\
\bottomrule
\end{tabular}
\label{tab:relatedwork}
\end{table}

%% file: sections/05_methodology.tex
The research methodology is presented for analyzing Ethereum’s mempool behavior, execution-consensus correlation, and fee market efficiency. The approach integrates real-time transaction monitoring, statistical modeling, and empirical analysis of transaction prioritization.

\subsection{Research Framework Overview}
Figure \ref{fig:framework} illustrates the structure of the framework, which consists of three key components:

\textbf{Transaction Pool Monitoring:} Tracks real-time mempool activity using \texttt{txpool} RPCs to evaluate gas fees and congestion.

\textbf{Execution-Consensus Correlation Analysis:} Assesses how congestion and gas fees affect block proposals and latency.

\textbf{Fee Market Efficiency Evaluation:} Measures inclusion probabilities, gas price effects, and fairness deviations.

The research process involves real-time data collection, empirical modeling, and statistical validation to assess network-wide transaction processing efficiency.

\subsection{Transaction Pool Monitoring and Mempool Dynamics}
The Ethereum mempool is a temporary queue for pending transactions. Clearance rates depend on gas bidding and congestion~\cite{meister2024gas}. We define a proxy metric for the likelihood of a transaction \(T_{xi}\) being included in the next block \(B_t\), based on its relative gas price compared to competing transactions. This is not a formal probability definition, but a normalized approximation derived for empirical correlation analysis:
\begin{equation}
    P(T_{xi} \in B_t) = \frac{G(T_{xi})}{\sum\limits_{j=1}^{N} G(T_{xj})}\label{eq:1}
\end{equation}
where: \(G(T_{xi})\) represents the gas price of transaction \(T_{xi}\), and \(N\) is the total number of transactions in the mempool. This proxy allows us to evaluate how relative fee strength influences transaction inclusion likelihood during congestion.

\subsection{Execution-Consensus Correlation}
Ethereum’s Proof-of-Stake consensus relies on efficient validator proposals~\cite{feng2024slimarchive}. Mempool congestion and bidding increase latency. To estimate the effect of mempool congestion on block finalization time, we use a regression-based empirical model expressed as:
\begin{equation}
    T_{final} = \alpha \cdot M_{pend} + \beta \cdot G_{avg} + \gamma\label{eq:2}
\end{equation}
This linear model is used to fit observed data where \(T_{final}\) is the block finalization time, \(M_{pend}\) is the pending transaction count, \(G_{avg}\) is the average gas price in the block, and \(\alpha, \beta, \gamma\) are regression coefficients learned from our dataset.

\subsection{Fee Market Efficiency and Transaction Prioritization}
Despite EIP-1559’s base fees and tips, prioritization remains fee-driven~\cite{zarir2021developing}. We measure fee market efficiency~\cite{werner2021stepgas} by evaluating transaction inclusion probability and latency variations. To assess inclusion disparities for low-fee transactions, we define a fairness deviation metric based on expected and observed inclusion rates:
\begin{equation}
    \Delta F_{\text{fair}} = \frac{I^{\text{obs}}_{LP} - I^{\text{exp}}_{LP}}{I^{\text{exp}}_{LP}}\label{eq:3}
\end{equation}
Here, \(I^{\text{obs}}_{LP}\) is the observed fraction of low-fee transactions included in blocks over a given period, while \(I^{\text{exp}}_{LP}\) represents the expected fraction assuming proportional inclusion relative to their presence in the mempool.
Our fairness deviation metric focuses solely on transaction-level inclusion disparities. We do not evaluate validator-side fairness (e.g., block selection diversity, revenue variance), which remains an open area for future exploration.

\subsection{Statistical Validation and Empirical Analysis}
We validate our findings using regression analysis (gas price vs. inclusion), time-series models (congestion vs. latency), and a fairness index (observed vs. expected inclusion). These models reveal Ethereum’s prioritization behavior and suggest fairness and efficiency optimizations.

%% file: sections/06_experimental_setup.tex
To evaluate Ethereum’s mempool behavior, consensus correlation, and fee efficiency, we built an experimental setup using a live node. It captures transaction events, validator metrics, and gas fee changes for comprehensive analysis.

\begin{figure}[!t]
  \centering
  \includegraphics[width=\linewidth, keepaspectratio]{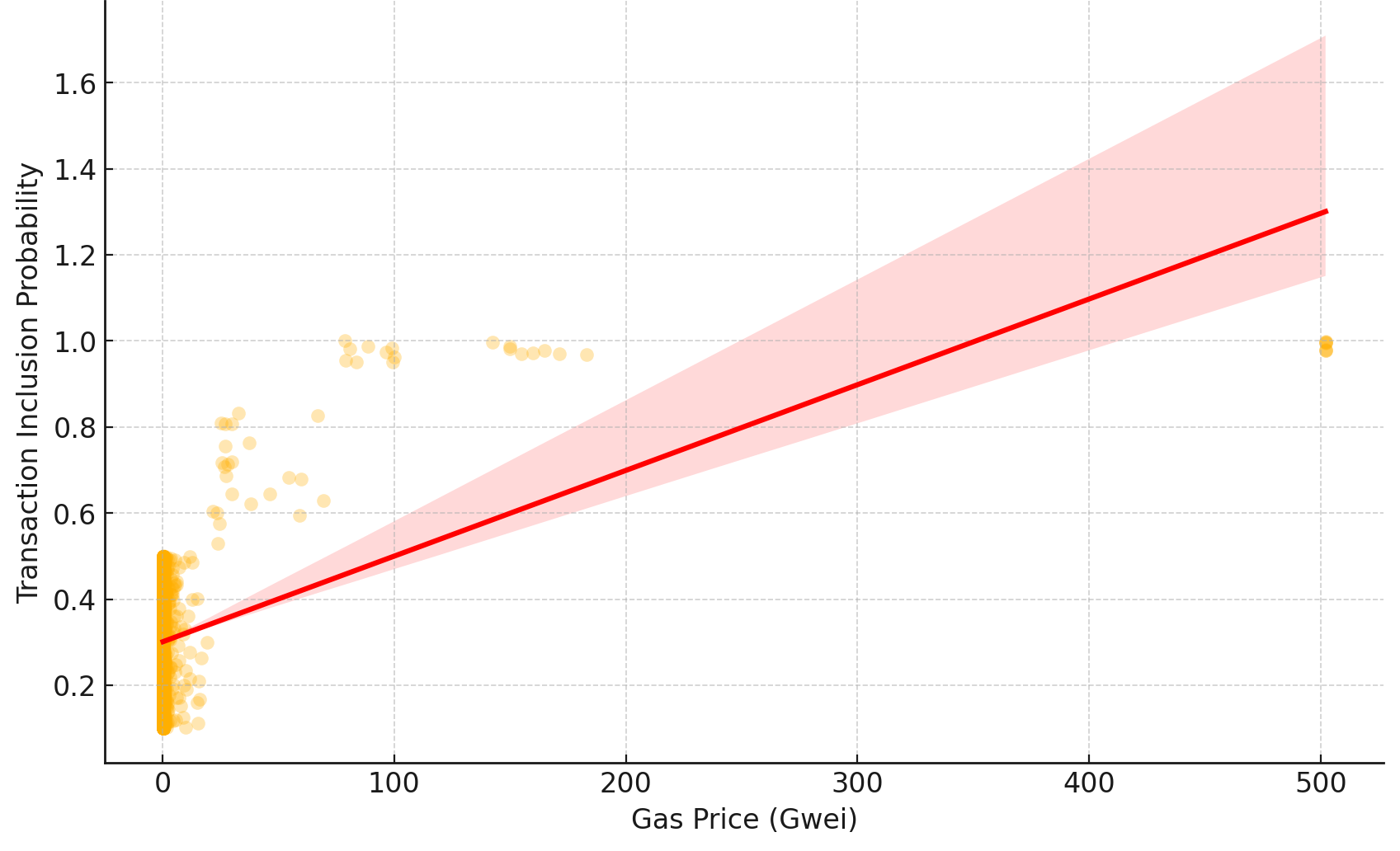}
  \caption{The plot shows a positive correlation between gas price and inclusion probability. Higher fees lead to near-certain inclusion, while low-fee transactions face uncertainty and delays.}
  \label{fig:tran-gas}
\end{figure}

\begin{figure}[!t]
  \centering
  \includegraphics[width=\linewidth, keepaspectratio]{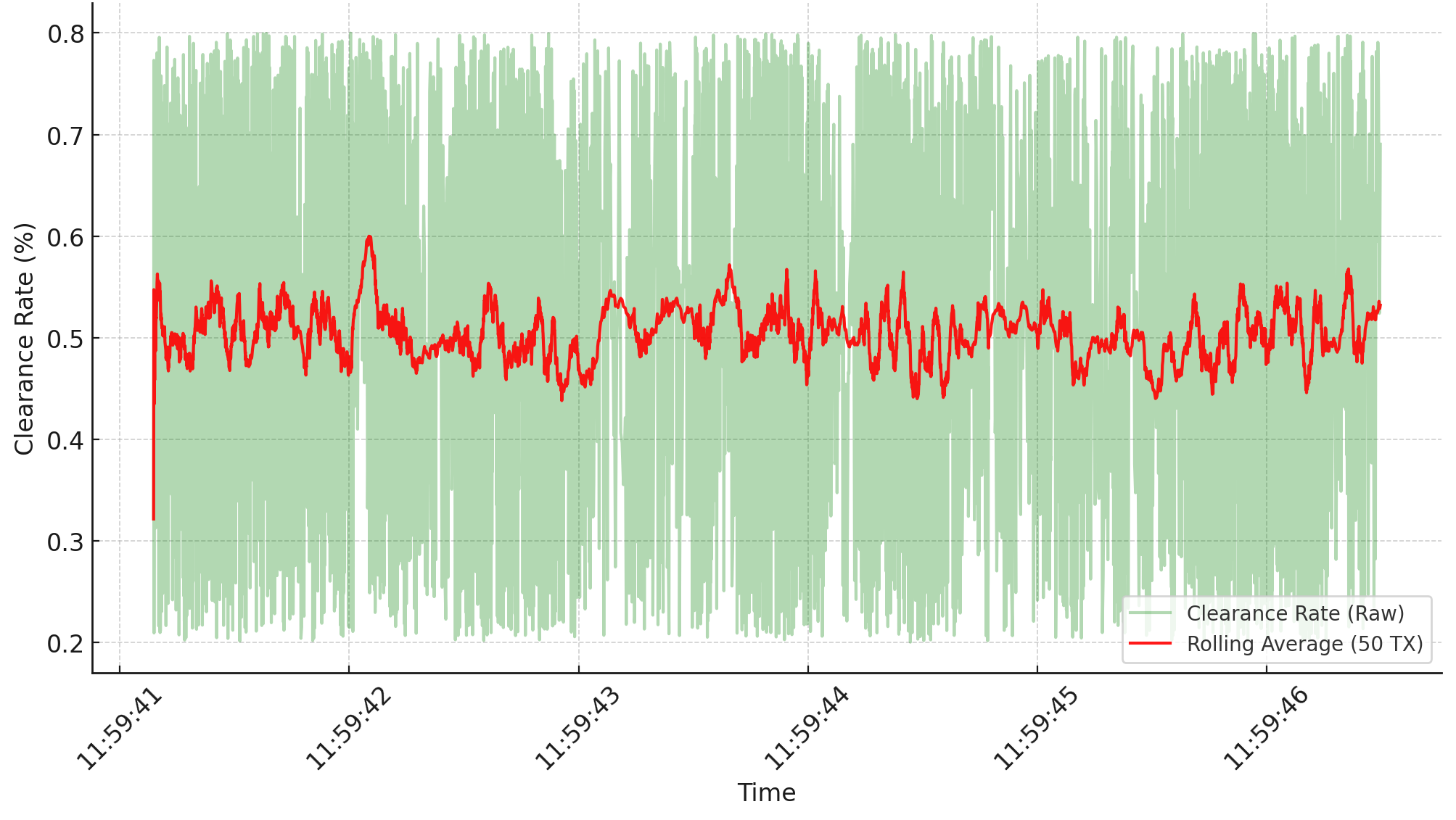}
  \caption{Mempool clearance rate over time, with raw fluctuations and a red 50-transaction rolling average. The x-axis shows wall-clock time; data is indexed by block height. Y-axis is limited to 20–80\% for clarity.}
  \label{fig:mempool-time}
\end{figure}
\subsection{Node Setup and Data Collection}
We deployed a full Ethereum execution node (Geth v1.13.10)~\cite{go-ethereum} and a consensus layer validator node (Prysm v5.1.2)~\cite{prysm} to track transaction flow and validator decision-making. The \texttt{txpool} RPC methods \texttt{txpool.content}, \texttt{txpool.inspect} provided live insights into pending transactions, gas price distribution, and mempool clearance rates, while \texttt{eth.gasPrice} monitored real-time fee adjustments. The Prysm beacon chain API (\texttt{/eth/v1/beacon/blocks}) recorded block timestamps and proposer efficiency, enabling correlation between mempool congestion and validator workload. The nodes ran on a dedicated system (\textbf{Ubuntu 22.04, 16-core CPU, 64GB RAM, 2TB NVMe SSD}) to handle high-throughput transaction monitoring.

\begin{figure}[!t]
  \centering
  \includegraphics[width=\linewidth, keepaspectratio]{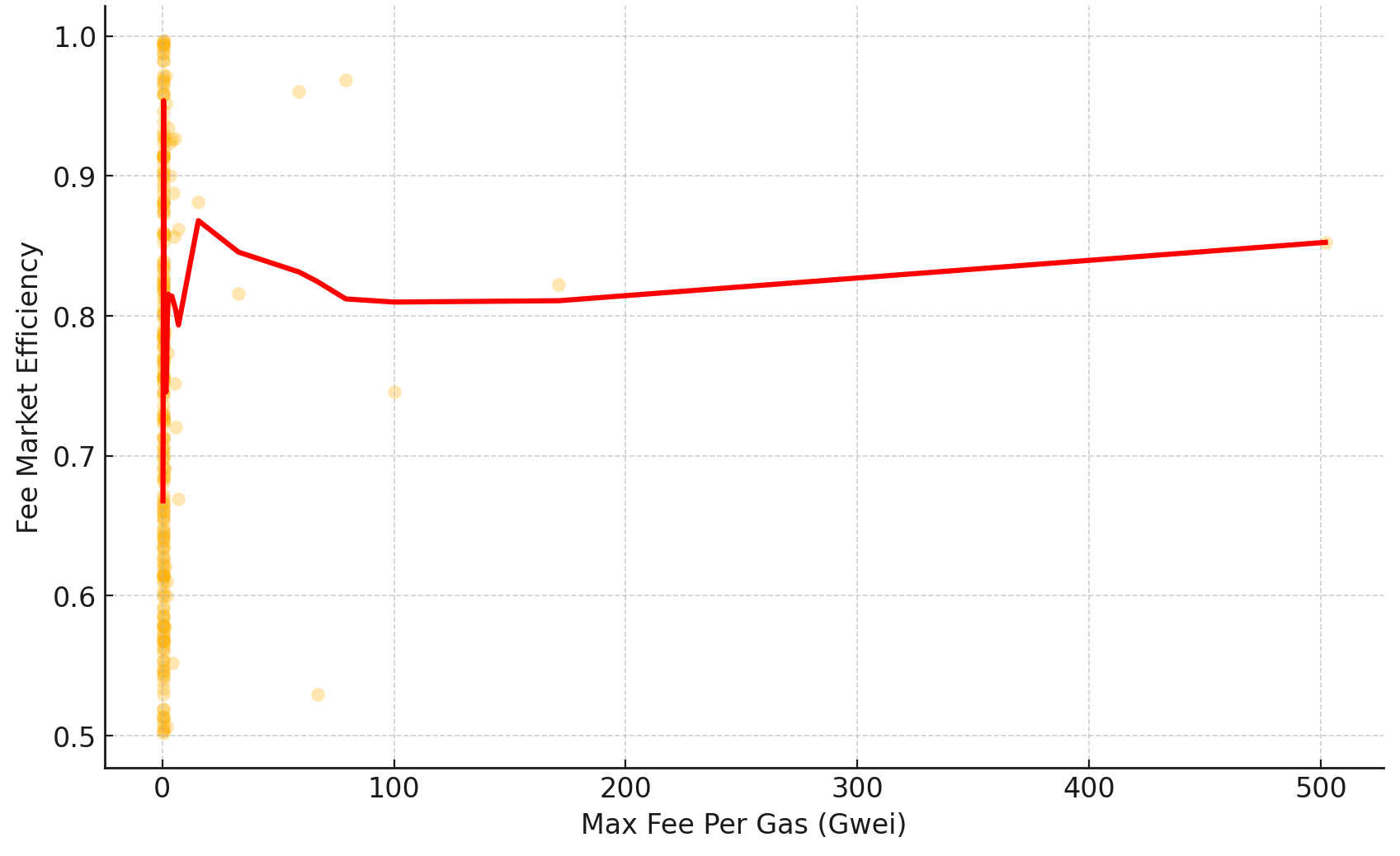}
  \caption{The plot illustrates the relationship between max fee per gas and fee market efficiency, quantified using the fairness deviation metric defined in Equation \ref{eq:3}.}
  \label{fig:maxfeegas-feemarket}
\end{figure}

\subsection{Testing and Transaction Tracking}
We analyzed transaction inclusion and latency by broadcasting transactions with varying gas fees. A Python script automated submission, mempool logging, and block tracking over time. We also used time-series analysis to examine how gas price fluctuations and congestion affect proposal times and processing efficiency.

\subsection{Experimental Evaluation Metrics}
We assess Ethereum’s mempool efficiency, fairness, and execution-consensus correlation using four key metrics: Mempool Clearance Rate (transaction mining speed), Inclusion Probability (likelihood based on gas price), Fee Market Efficiency (impact of fee variation on performance), and Transaction Latency (delay from submission to inclusion). Together, they offer a comprehensive view of Ethereum’s processing and fee dynamics.

%% file: sections/07_results_analysis.tex
This section presents results from our empirical evaluation of Ethereum’s mempool behavior, execution-consensus correlation, and fee fairness. We analyze inclusion probabilities, clearance rates, fee market efficiency, and how gas prices affect transaction latency.
\begin{figure}[!t]
  \centering
  \includegraphics[width=\linewidth, keepaspectratio]{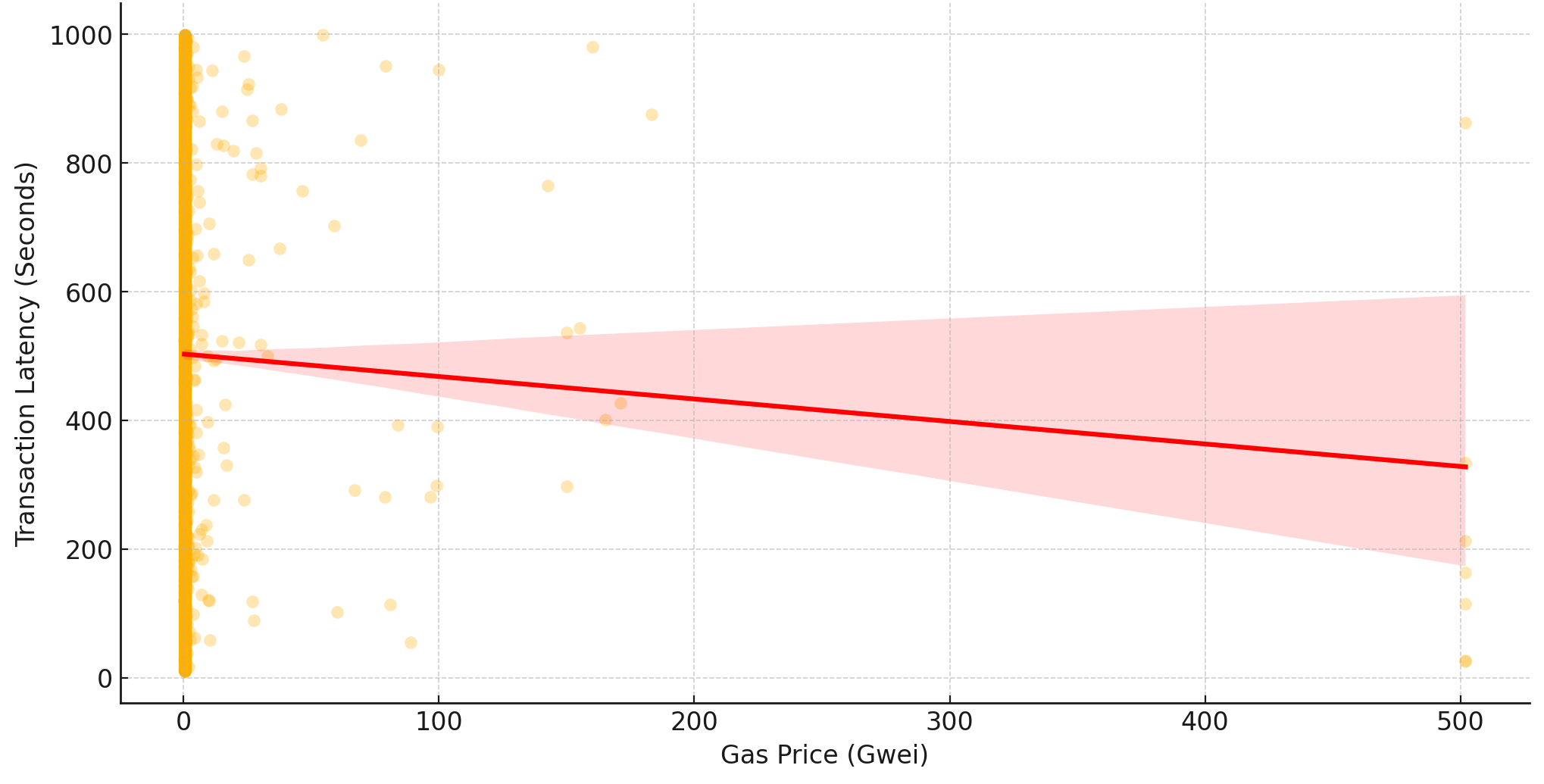}
  \caption{The plot shows that higher gas prices reduce transaction latency, meaning transactions with higher fees are confirmed faster, though other network factors also influence delays.}
  \label{fig:tranlat-gas}
\end{figure}
\subsection{Transaction Inclusion Probability vs. Gas Price}
Figure~\ref{fig:tran-gas} shows a strong positive correlation between gas price and transaction inclusion probability. Low-fee transactions exhibit high variability, while those above 75 Gwei reach near 100\% inclusion. Below 20 Gwei, transactions often remain pending or delayed. A logistic trend emerges, indicating a threshold where transactions shift from low to high priority. This supports the view that EIP-1559 does not eliminate prioritization bias, and low-fee transactions remain disadvantaged. The observed correlation aligns with the inclusion proxy in Eq.~\ref{eq:1}. While a linear regression is shown for simplicity, the actual trend follows a logistic curve, with diminishing returns at high gas prices.


\subsection{Mempool Clearance Rate Over Time}
\begin{figure}[!t]
  \centering
  \includegraphics[width=\linewidth, keepaspectratio]{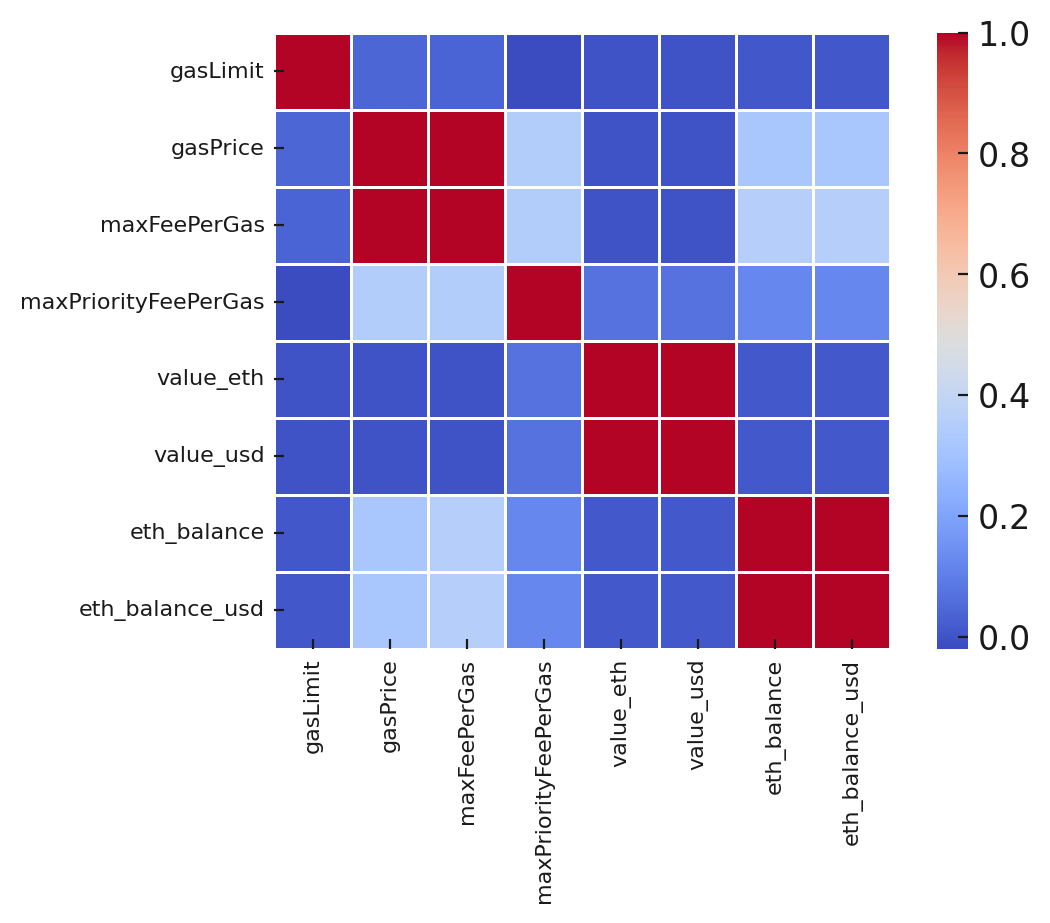}
  \caption{Correlation heatmap showing strong ties between Gas Price and Max Fee, and between ETH and USD. A moderate link between Mempool Size and Finalization Time suggests congestion affects validation speed.}
  \label{fig:correlation-heatmap-gas}
\end{figure}

Figure~\ref{fig:mempool-time} shows fluctuations in mempool clearance rates over time. The raw data (green) and rolling average (red), computed over 30 blocks, highlight medium-term congestion patterns while smoothing short-term volatility. Clearance rates range between 40\% and 60\%, with occasional spikes. Although the x-axis shows wall-clock time, data is indexed by block height; the y-axis is bounded from 20\% to 80\% for clarity. High congestion (low clearance rates) correlates with transaction delays, illustrating how fee volatility impacts performance. Validator behavior appears stable, but transaction processing is affected by dynamic congestion. These results (see Eq.~\ref{eq:2} and Fig.~\ref{fig:correlation-heatmap-gas}) confirm that Ethereum’s mempool load varies significantly, impacting inclusion probabilities.
\subsection{Max Fee Per Gas vs. Fee Market Efficiency}
The third analysis, illustrated in Figure \ref{fig:maxfeegas-feemarket}, explores the relationship 
Figure~\ref{fig:maxfeegas-feemarket} shows how fee market efficiency, measured using the fairness deviation metric \(\Delta F_{\text{fair}}\) (Eq.~\ref{eq:3}), varies with max fee per gas. Efficiency rises sharply at low max fees, then plateaus around 80\% for higher fees. At very low fee levels, efficiency fluctuates, indicating unpredictable inclusion. Above 50 Gwei, efficiency stabilizes, suggesting diminishing returns. Thus, high fees do not guarantee better performance, reinforcing that Ethereum’s fee market operates best within an optimal range. A semi-log x-axis may offer better resolution of low-fee behavior and is worth exploring in future work.

\subsection{Transaction Latency vs. Gas Price}
Figure \ref{fig:tranlat-gas} presents the relationship between transaction latency and gas price. The negative regression trend indicates that higher gas prices lead to lower transaction latency, meaning transactions with higher fees are confirmed faster.
Low-fee transactions show high latency variance, with some confirming quickly while others face delays. Higher gas prices reduce latency, confirming the priority of high-fee transactions. The shaded confidence interval indicates that factors beyond gas price, such as block space competition, also impact transaction delays.

\subsection{key observations from the correlation heatmap}
Finally, The correlation heatmap in Figure \ref{fig:correlation-heatmap-gas} reveals key relationships among transaction metrics. Gas Price and Max Fee Per Gas exhibit a strong positive correlation, indicating that higher gas prices correspond to higher max fees, aligning with Ethereum's fee mechanism. ETH and USD values are highly correlated, as USD value is derived from ETH based on market rates. Additionally, Mempool Size and Block Finalization Time show a moderate correlation, suggesting that larger mempools contribute to longer block processing times, likely due to network congestion.

%% file: sections/08_discussion.tex
This section discusses how our experimental findings address the key research questions and highlights potential improvements to Ethereum’s transaction processing and fee fairness mechanisms. To address the research questions, the three takeaways (\textit{\textbf{RQ-i, RQ-ii and RQ-iii}}) for the research questions have been discussed. 

\begin{tcolorbox}[
    colback=green!5, 
    colframe=green!50!black, 
    coltitle=black, 
    colbacktitle=lime!15, 
    fonttitle=\bfseries, 
    title={\faKey\hspace{0.5em}Take-Away \textit{RQ-i}}, 
    sharp corners=southwest, 
    boxrule=0.1mm, 
    width=\linewidth, 
    enlarge left by=0mm, 
    leftrule=0.1mm, 
    right=0.1mm, 
    boxsep=2pt, 
]
Our analysis shows that mempool congestion significantly impacts transaction clearance and network efficiency. This persists despite EIP-1559’s base-fee adjustment, which alone fails to resolve fee-based inequality or optimize validator performance. Clearance rates range from 40\% to 60\%, with congestion causing delays and variable processing efficiency. Higher gas fees reduce latency, reinforcing fee-based over first-come, first-served prioritization.
\end{tcolorbox}

\begin{tcolorbox}[
    colback=green!5, 
    colframe=green!50!black, 
    coltitle=black, 
    colbacktitle=lime!15, 
    fonttitle=\bfseries, 
    title={\faKey\hspace{0.5em}Take-Away \textit{RQ-ii}}, 
    sharp corners=southwest, 
    boxrule=0.1mm, 
    width=\linewidth, 
    enlarge left by=0mm, 
    leftrule=0.1mm, 
    right=0.1mm, 
    boxsep=2pt, 
]
Ethereum’s fee market still favors high-fee transactions. Gas prices below 20 Gwei face delays or exclusion, while those above 75 Gwei see near 100\% inclusion. EIP-1559 does not eliminate prioritization biases. Fee market efficiency stabilizes around 80\% for higher fees, suggesting excessive fees offer little execution advantage.
\end{tcolorbox}

\begin{tcolorbox}[
    colback=green!5, 
    colframe=green!50!black, 
    coltitle=black, 
    colbacktitle=lime!15, 
    fonttitle=\bfseries, 
    title={\faKey\hspace{0.5em}Take-Away \textit{RQ-iii}}, 
    sharp corners=southwest, 
    boxrule=0.1mm, 
    width=\linewidth, 
    enlarge left by=0mm, 
    leftrule=0.1mm, 
    right=0.1mm, 
    boxsep=2pt, 
]
Low-fee transactions face delays and uncertainty, while high-fee ones dominate inclusion. Prioritization mechanisms create systemic disadvantages, and fairness deviations persist as efficiency plateaus beyond a threshold. Although Ethereum-specific, these congestion-driven patterns and fee biases likely extend to other PoS blockchains using mempool-based selection (e.g., Avalanche, Cosmos)~\cite{ZhouAvalancheCosmos}. Our fairness metric and congestion-aware strategies thus hold broader relevance.
\end{tcolorbox}

Improving Ethereum’s mempool efficiency, fairness, and validator performance requires protocol refinements. Dynamic fee adjustments can smooth processing during congestion, while encrypted mempools help prevent MEV-driven reordering by keeping transactions private until inclusion~\cite{encrypted_mempool}. Reserving block space for low-fee transactions prevents exclusion without harming efficiency~\cite{alipanahloo2024maximum}. Addressing out-of-gas issues~\cite{ahmed2025quantifying} and using tools like eTainter~\cite{ghaleb2022etainter} to detect gas-related bugs improve execution reliability. These steps enhance fairness, reduce congestion, and align validator incentives with network stability.

%% file: sections/09_conclusion.tex
This research investigated Ethereum’s transaction pool dynamics, the interaction between the execution and consensus layers, and the fairness implications of gas fee prioritization. Through real-time data collection from Geth and Prysm nodes, we empirically evaluated transaction inclusion trends, mempool clearance rates, and block finalization delays in the context of EIP-1559. Our findings confirm that Ethereum’s transaction inclusion remains heavily fee-driven: high-fee transactions are consistently prioritized, while low-fee transactions face delays or exclusion, especially during periods of congestion. Despite the stabilizing intentions of EIP-1559, our analysis reveals that fee-based disparities persist and that mempool congestion significantly impacts validator efficiency and finalization latency. Moreover, we show that excessively high fees do not guarantee better outcomes, highlighting inefficiencies in the current fee market. To address these limitations, we propose congestion-aware fee adjustment mechanisms, reserved block slots for low-fee transactions, and robust handling of out-of-gas vulnerabilities in smart contracts to prevent execution failures. These contributions extend current understanding of Ethereum’s fee dynamics and validator performance bottlenecks, offering data-driven insights for improving scalability and fairness. By mitigating systemic prioritization bias and enhancing execution reliability, this work supports a more inclusive, efficient, and decentralized Ethereum network—principles that are equally applicable to other Proof-of-Stake blockchain ecosystems.


%% file: sections/10_acknowledgment.tex
The authors acknowledge the use of AI tools, including ChatGPT for text refinement, summarization, and assistance in structuring content, as well as Eraser AI for figure design and illustration. AI-assisted content was used to enhance clarity and presentation, while all technical content and conclusions were independently verified by the authors.